# A near-infrared SETI experiment: instrument overview

Shelley A. Wright[*a,b], Dan Werthimer[c], Richard R. Treffers[d], Jérôme Maire[a], Geoffrey W. Marcy[e], Remington P.S. Stone[f], Frank Drake[g], Elliot Meyer[b], Patrick Dorval[b], Andrew Siemion[e]

[a]Dunlap Institute for Astronomy & Astrophysics, Univ. of Toronto, Toronto ON, M5S 3H4, Canada,
[b]Dept. of Astronomy & Astrophysics, Univ. of Toronto, Toronto, ON, M5S 3H4, Canada;
[c]Space Sciences Laboratory, University of California Berkeley, Berkeley, CA 94720, USA;
[d]Starman Systems, LLC, Alamo, CA 94507, USA;
[e]Astronomy Department, University of California Berkeley, Berkeley, CA 94720, USA;
[f]Lick Observatory, University of California Santa Cruz, Santa Cruz, CA 95064 USA;
[g]SETI Institute, Mountain View, CA 94043 USA

## ABSTRACT

We are designing and constructing a new SETI (Search for Extraterrestrial Intelligence) instrument to search for direct evidence of interstellar communications via pulsed laser signals at near-infrared wavelengths. The new instrument design builds upon our past optical SETI experiences, and is the first step toward a new, more versatile and sophisticated generation of very fast optical and near-infrared pulse search devices. We present our instrumental design by giving an overview of the opto-mechanical design, detector selection and characterization, signal processing, and integration procedure. This project makes use of near-infrared (950 – 1650 nm) discrete amplification Avalanche Photodiodes (APD) that have > 1 GHz bandwidths with low noise characteristics and moderate gain (~$10^4$). We have investigated the use of single versus multiple detectors in our instrument (see Maire et al., this conference), and have optimized the system to have both high sensitivity and low false coincidence rates. Our design is optimized for use behind a 1m telescope and includes an optical camera for acquisition and guiding. A goal is to make our instrument relatively economical and easy to duplicate. We describe our observational setup and our initial search strategies for SETI targets, and for potential interesting compact astrophysical objects.

## 1. INTRODUCTION

The Search for Extraterrestrial Intelligence (SETI) has undertaken a variety of searches for signals or communications from extraterrestrial intelligence (Drake[1], Tarter et al.[2]). There has been growing interest in searches for fast (nanosecond) optical pulses emanating from transmitting beacons from extraterrestrial intelligence (e.g., Horowitz et al.[3], Werthimer et al.[4], Wright et al.[5], Reines & Marcy[6], Howard et al.[7], Stone et al.[8], Howard et al.[9], Covault[10]). Today's highest-powered pulsed lasers can outshine our Sun by at least 4 orders of magnitude in brightness, and may offer one of the best means for interstellar communication (as originally suggested by Schwartz & Townes[11]). Such lasers have been developed for a number of applications on Earth and are well within our current technological capabilities. For instance, research groups at Lawrence Livermore National Laboratory have generated laser pulses with peak power of the order of a terawatt for times lasting many picoseconds (Moses[12,13]). Such pulses, when concentrated into a narrow beam by a large telescope such as the current 8-10m reflectors, create a photon flux in their beams so powerful that they easily outshine all the light of the host star. These high power pulses would be detectable with meter-class telescopes even across great distances (~1,000s light years) within the Milky Way. A great advantage of these lasers is that they can be narrowly beamed, thus providing the highest amount of transmitted power flux and information per unit energy. Therefore, using such a laser offers a promising means for interstellar communication, and it is possible that an advanced extraterrestrial civilization would choose to communicate via pulsed lasers. In order to be detected, such interstellar signals must be distinguishable from other background signals, including terrestrial, atmospheric, and astrophysical phenomena. While astrophysical objects that produce nanosecond or shorter optical pulses are very rare (Howard et al.[7]), this is an area requiring further investigation at optical and near-infrared wavelengths.

---

[*] Send correspondences to wright @ astro.utoronto.ca

Near-infrared (1000 - 3500 nm) astronomy has matured rapidly in the last decade, with more advanced infrared detectors offering higher quantum efficiencies and lower detector noise. At the same time, near-infrared lasers and detectors are being developed and explored for a variety of applications, in particular within the telecommunications industry. One major advantage for interstellar communication of using longer wavelengths is the decrease in interstellar extinction (Cardelli et al.[14]), which is of particular importance for communicating close to the plane of the Milky Way. For instance, at visual wavelengths (~600 nm) there are 30 magnitudes of extinction looking towards the Galactic Center, whereas at IR wavelengths (~1600 nm) there are only 2 magnitudes of extinction (Nishiyama et al.[15]). This is particularly highlighted by examining the extinction law ($A_\lambda/A_V$; Cardelli et al.[14]) through the Galactic plane (Figure 1); at 1 kpc distance with a 1.5 micron source the interstellar medium is 65% transmissive, whereas at 0.5 micron it is less than 10%. In addition, Galactic background from warm dust peaks at wavelengths in the mid- to far-infrared, beyond the near-infrared. Therefore, near-infrared offers a unique window with less interstellar extinction and less background from our galaxy, where signals can be efficiently transmitted at larger distances within the plane of the galaxy. This near-infrared regime has already been identified as an optimum spectral region for interstellar communications (Townes[16]), yet this has remained unexplored territory for SETI.

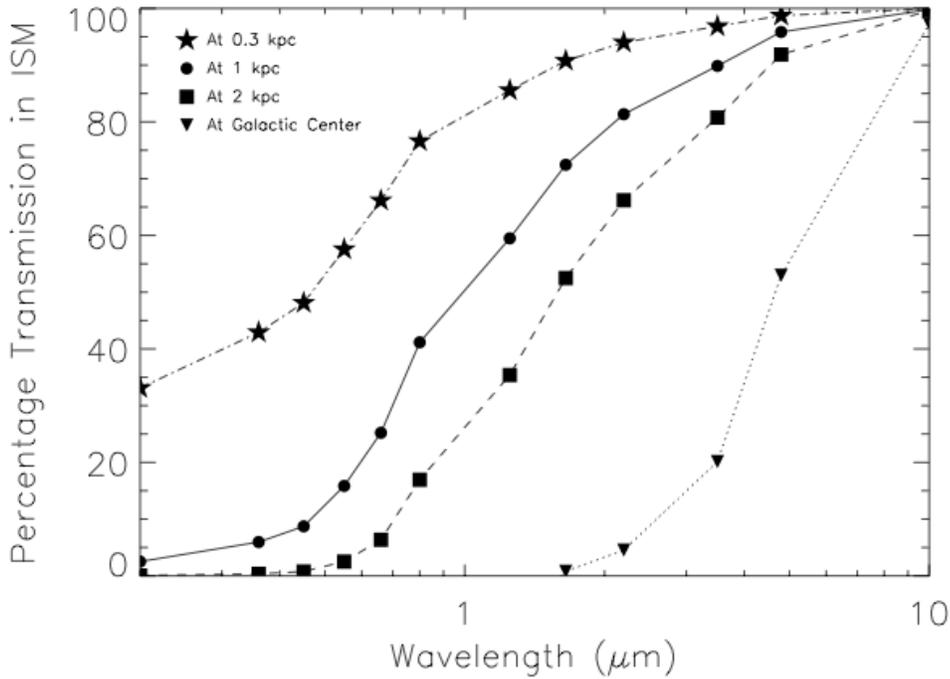

**Figure 1**: Percentage of flux transmitted through the interstellar medium (ISM) at fixed distances through the Galactic plane for wavelengths from 0.2 to 10 μm. These curves were generated using the extinction law from Cardelli et al.[14] and assuming measured extinction values averaged through the Galactic Plane, except when towards the Galactic Center. We show the percentage of transmission at 0.3 kpc or ~1,000 light years (star points, dot-dashed line), 1 kpc (circle, solid line), 2 kpc (squares, dashed line), and at the Galactic Center (triangles, dotted line).

Fast-response near-infrared time-domain astrophysics (<1 μs) is an unexplored phase space. Short pulses from a variety of compact celestial objects such as neutron stars, novae, supernovae, black holes, and active galactic nuclei (AGN) at near-infrared wavelengths may prove interesting. Suggestions that such searches could bear important fruit come from, for example, the fact that radio pulses observed in some pulsars are unresolved in time (e.g., Hankins &

Boriakoff[17], Kuzmin et al.[18]), and shorter in duration than a microsecond. This implies very high energy densities exist in the emitting regions, in turn suggesting the possibility of associated optical radiation. Similarly, "giant" radio pulses from the Crab Nebula are unresolved down to nanosecond intervals (Hankins et al.[19], Cordes et al.[20]), and have been recently attempted to be resolved in the optical domain with fast photometers (Germanà et al.[21]). New transient surveys (e.g., Palomar Transient Factory: Law et al.[22]) are still discovering new classes of compact objects and supernovae, and finding new optical transients on ever-shorter timescales (e.g., Cenko et al.[23]). There is already interest in using fast-response optical and near-infrared detectors for future astronomical quantum optics experiments for the extremely large telescopes, see Dravins et al.[24]. Fast photonic astronomy is hoped to be a new field with these behemoth telescopes; they would like to achieve down to picosecond resolutions to begin to detect the coherence of light from astronomical objects. There are already efforts underway to reach down to micro-second and nano-second precision at optical wavelengths. For instance, new optical photometers have begun searches around pulsars to resolve nano-second structure (Naletto et al.[25]). Therefore searches for pulsed laser signals from extraterrestrial intelligence have the distinct advantage of potential for unveiling exciting new areas of time-domain astrophysics and pushing the instrumentation in this field.

We describe our efforts to design a new innovative near-infrared instrument to search for nanosecond pulses emanating from an extraterrestrial civilization. We will search not only for transient phenomena from technological activity, but also from natural objects that might produce very short time scale "flashes" such as pulsars, AGN's, black holes, gamma ray bursts, and other esoteric objects. We will also enhance our instrument with the ability to record detailed time behavior of any flashes detected, and will provide post-processing algorithms for searching within the data. In Section 2 we review the instrument with the detector selection and characterization (Section 2.2) and optical and mechanical layout (Section 2.3). We note that the detector characterization and analysis techniques are presented by Maire et al., this conference (9147-173). In Section 3, we discuss our commissioning plans and integration at the telescope, and initial observational plans. We summarize our instrument in Section 4.

## 2. INSTRUMENT OVERVIEW

### 2.1 Conceptual design of "optical pulsed" SETI

Early programs to detect nanosecond optical flashes used single detectors, which were greatly troubled by false positive signals. These distracting internal events can be produced in several ways, including radioactive decay in the envelopes of the photomultiplier tubes (PMT's) universally used for detection of brief optical pulses, sparks from corona discharge, or by cosmic ray hits. For our original detection system, a very successful means to avoid such false positives was developed by Werthimer and Horowitz. The method depends on dividing the captured light into multiple streams, the photons of each stream being delivered to a separate detector. The photon stream are carried along optical and then electrical paths of carefully controlled equal length to a coincidence board which will be triggered if all the pulses arrive simultaneously. Thus, spurious photon events produced within individual detectors are ignored. A coincidence will only occur if there is strong deviation from Poisson statistics, which is the trademark of a simultaneous burst of many synchronized photons from a laser. Such a deviation from Poisson statistics can give strong suggestion of technological origin of the photons. For the optical SETI experiment at both Leuschner and Lick Observatory that use PMTs, it was determined that the optimal number of detectors that had good sensitivity and low false alarm rate was three (Werthimer et al.[4], Wright et al.[5]). For our near-infrared experiment using significantly different detectors, we re-explored the probability distribution of false alarm rate with the new near-infrared detectors, discussed in Maire et al., this conference.

### 2.2 Selection of near-infrared detectors

Our team has been investigating the capabilities of fast-response (> 1 GHz) near-infrared detectors with sensitivity to detect single photons. Near-infrared PMTs traditionally have low quantum efficiency (QE) of 1-2%, which quickly led us to look into the rapidly evolving near-infrared APDs. To extend to near-infrared wavelengths (900 – 1700 nm) with our required time resolution, we investigated InGaAS APDs. We felt this wavelength coverage was a good start since we wanted to keep the project economical and to avoid additional complexity and expense of cryogenics.

We first explored APDs that operated in either linear or Geiger mode. There were distinct advantages for either readout version: Geiger mode APDs offer high gains (M $>10^{4-5}$), but have large quenching times of order of 0.1 – 100 microseconds (Williams et al.[26]). In contrast, linear APDs have low quenching times (< 10 nanoseconds) but very low gains (M < 50). Another important design consideration was size of the active area of the APDs; for the optical design we wanted a large active area for ease of alignment and adequate field of view on-sky. Our team investigated many vendors and models of APDs for Geiger and linear modes and determined that neither was satisfactory, due to available trade-offs (excess noise factor, noise equivalent power, QE and dead-time).

Given these considerations, this led our team into investigating relatively new APDs using an internal discrete amplification[27,28] readout process (Krutov et al.[29], Linga et al.[30]). These discrete amplification APDs offer lower quenching times (<10-15 ns) and allow flexibility in the choice of readout given the application. Discrete amplification makes use of an array of micron-size photodetectors or "pixels" that have individually controlled amplification; each of these signals are then combined to provide a single analog output. The use of multiple "pixels" means that these composite APDs can be operated in the Geiger mode with high gains at the individual detector level, while maintaining low quench times (< 10 ns) for the device as a whole. Thus, discrete amplification APDs can detect faint and fast (~ nanoseconds) pulsed signals, while also maintaining low noise and large dynamic range. To test the performance of the latest discrete amplification APD we purchased 80 μm detectors from Amplification Technologies[*]. We conducted several months of lab testing to determine the best operating biases for our program, considering dark count rates and pulse height distributions. These measurements were important for determining false alarm rates given the selected number of detectors in the instrument. These tests and further discussion of pulse height distributions and false alarm rates are presented by Maire et al., this conference.

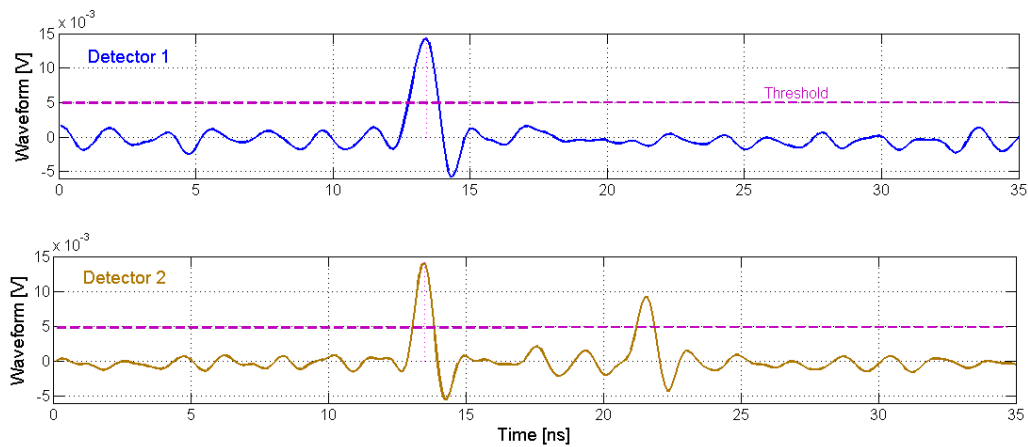

**Figure 2**: Two discrete amplification APD waveforms showing a pulsed signal from a 1.55 μm pulsed laser source. A beam-splitter divides the light path into the two detectors, where a pulse arrives simultaneously at 13 ns. A secondary dark pulse is seen in detector 2 at 22ns. The detection threshold can be adjusted to trigger on any signal above $5\times10^{-3}$ V in the above experiment.

From our detector testing, we determined that we should be able to operate our instrument in either a single or two detector mode; the choice of the number of detectors is dependent on the photon rate and the threshold and therefore the sensitivity achieved. The threshold per detector can be adjusted based on dark count rate and photon count rate per source to trigger any signal above a certain voltage, as illustrated in Figure 2. All signal processing is performed with a high-bandwidth oscilloscope. We currently use an oscilloscope with 2.5GHz bandwidth, 20 giga-samples per second, and a standard memory of 20 million points per channel. We have written software that allows choice of triggering criteria, and recording of any portion of the waveform for post analysis, if deemed interesting. Another important benefit of this new system will be the introduction of a very flexible digital back-end that will greatly increase the capabilities. We can then examine these recordings to help understand post-facto the nature of any detection, or perform more advanced analysis for either SETI or astrophysical study.

The 80 μm APD is housed in a TO-8 can that is cooled by a thermal electric cooler (TEC). Loosely based on Amplification Technologies evaluation modules, we designed a readout board that allows control of operating bias, TEC, monitors temperature on the back-side of the TO-8 can, and analog output signal. This output signal is then fed into a pre-amplified and then digitized by an oscilloscope. Typical operating bias for a cooled 80μm APD at -30C is 58 – 65 V. Current across the APD has to be essentially zero, with a power supply that needs a compliance current of less than 10 μA. We designed a power supply board that allows the required adjustable voltage with a low compliance current. Figure 3 illustrates both the readout board for the TO-8 can and power supply board.

---

[*] Amplification Technologies Inc., 50 Eisenhower Drive, Paramus, NJ 07652, USA

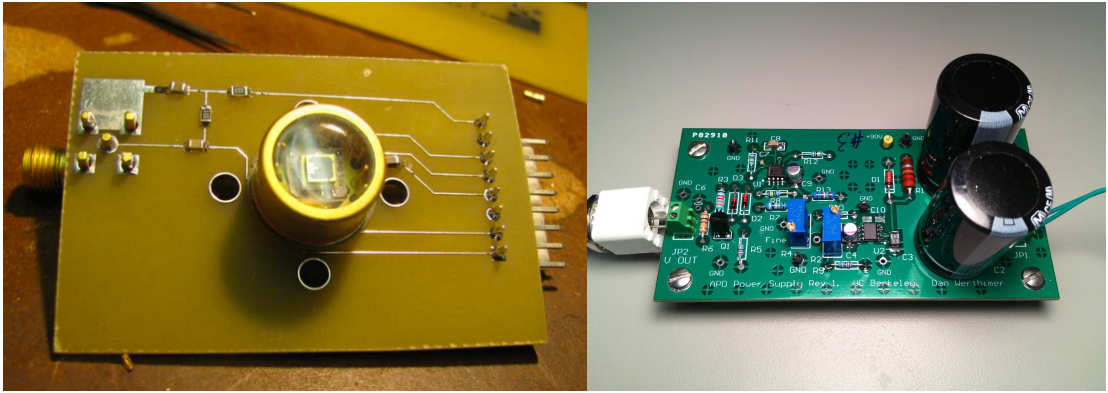

**Figure 3**: (LEFT) TO-8 can with the 80 m discrete amplification APD inserted in our in-house readout and control board, by R. Treffers. (RIGHT) Custom-made power supply board that delivers operating bias to an individual APD with low current (< 10 μA) and tunable voltage up to 80V, designed by D. Werthimer.

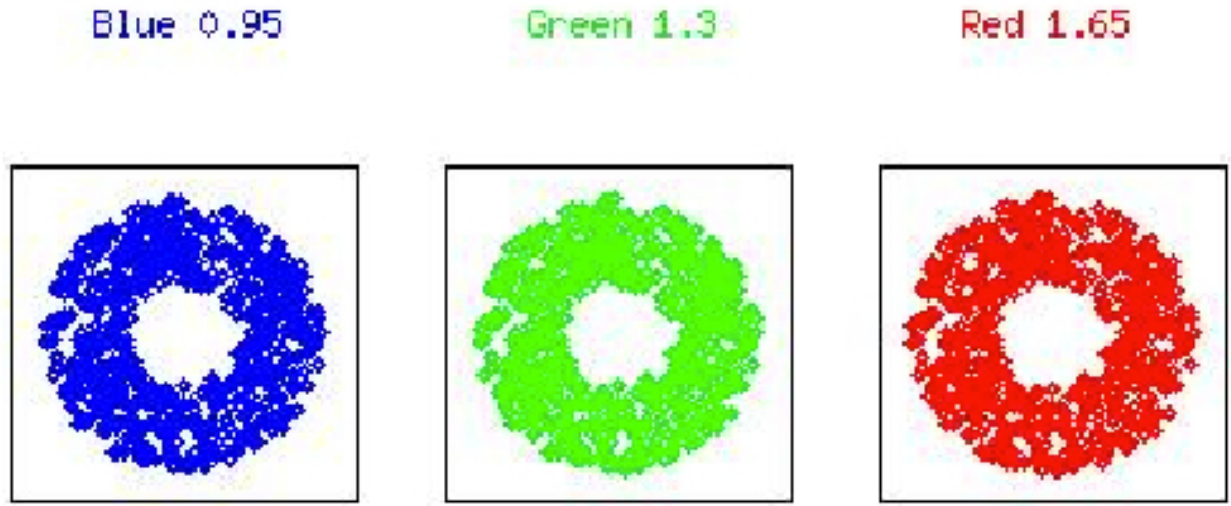

**Figure 4**: Spot diagram across the 80 um detector diameter (black square) at three wavelengths: 0.95, 1.3, and 1.65 um. We have selected the optical design to evenly illuminate the APD with a sharp cutoff in the light distribution across each detector.

## 2.3 Optical and Mechanical layout

Our initial plans are to install the instrument at Lick Observatory on the 1m Nickel Telescope with an f/17 input. Our optical design was optimized to achieve the largest field of view, while also illuminating the detector uniformly. The design is a two-detector system that has a single 50/50 beamsplitter. Since the detectors are only 80μm in size the field of view using the Nickel f/ratio would only be 1" in diameter. Therefore optics must be used to increase the FOV on-sky to allow for seeing variations and guiding errors. Use of a single lens with a long focal length followed by a beam splitter would have aberrations too large to allow a clean split between two detectors. Since these APDs are multiple photodiodes, the light needs to be spread uniformly across them. We select a double lens system with a -15mm focal length diverging lens placed before the beam splitter, and a 15mm focal length field lens placed just before the detector.

The combination of diverging and field lenses yields an f/ratio magnification of 15.1 across the detector with an even illumination, yielding a FOV of 4.4" on-sky. A spot diagram is shown in Figure 4 with a star centered in the FOV, which shows a well-illuminated surface across the 80 μm detector across our bandpass.

All optical and mechanical components were selected to be off the shelf from well-established commercial vendors, like Thorlabs and Edmund Optics. The detectors with their associated readout boards are installed in a C-mount and are each housed in a kinematic mount that can be adjusted for initial alignment. The f/17 input from the telescope enters a baffled aluminum instrument housing. The first optical element is a dichroic that splits the optical wavelengths (400 – 680 nm) into the guide camera, and the near-infrared wavelengths (720 - 1700nm) into the science detectors. Following the dichroic, a lens tube houses the diverging lens, followed by a filter wheel with selectable near-infrared astronomical filters installed (e.g., J and H band) and neutral density filters for very bright stars. After the filter wheel, a translation stage may be inserted with a 50/50 beamsplitter to select between using one or two detectors. The field lenses are installed on the kinematic mount in front of the detectors. We have also installed a near-infrared 1.55 μm laser that is attached to pulse driver that can generate at 10 ns width pulses. This is used for system verification and calibration of the APDs. The opto-mechanical design is presented in Figure 5. A cartoon schematic of the system illustrates the light path and components of the instrument in Figure 6.

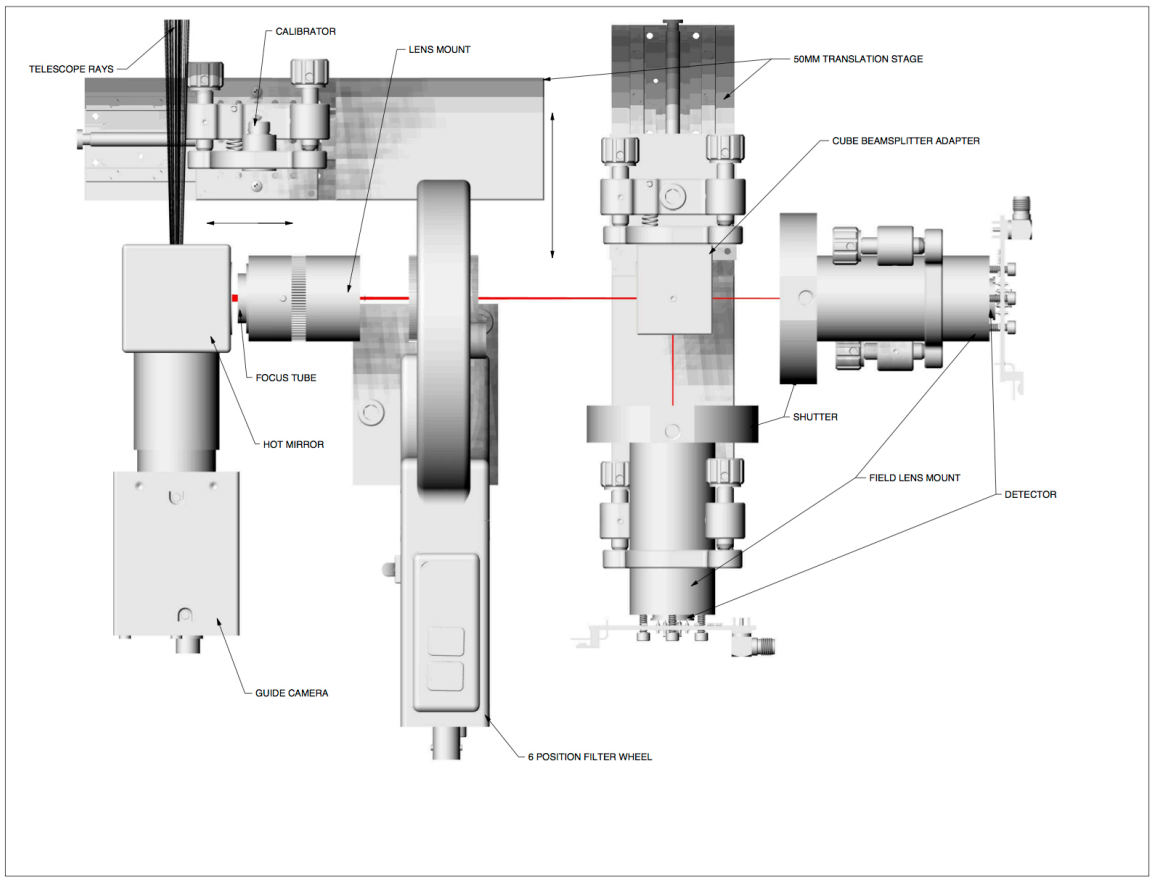

**Figure 5:** The optical-mechanical layout of the NIROSETI instrument. The light path from the telescope arrives at the top-left and immediately enters a dichroic that splits the light into optical wavelengths for the guide camera and near-infrared wavelength into the science path. After a diverging lens followed by a filter wheel the light path can then can go either into single detector, or if a beamsplitter is inserted into the light path with a translation stage, the light can be split 50/50 between two detectors. Both detectors are housed in C-mounts attached to kinematic mounts with a field lens in front of each detector.

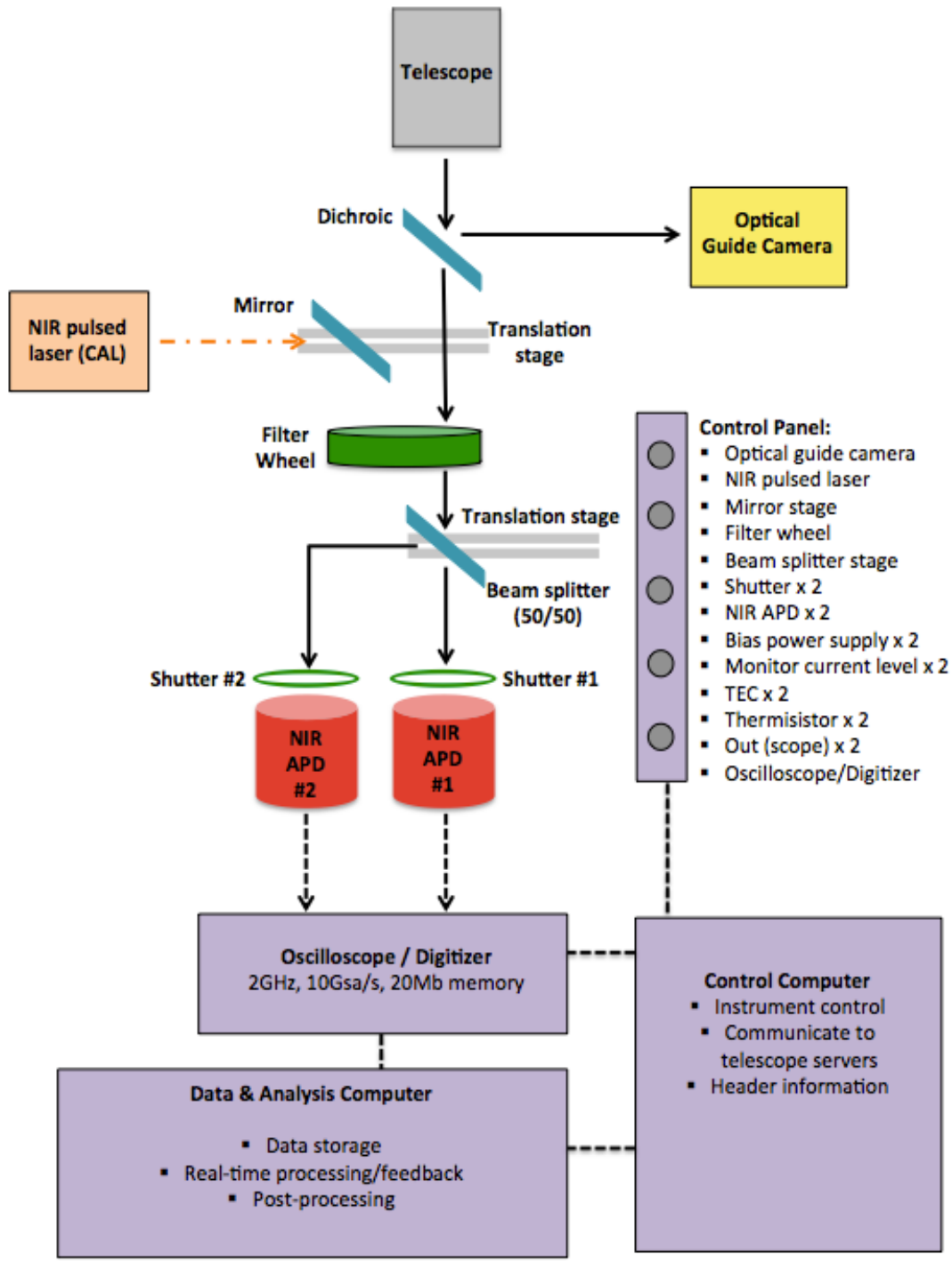

**Figure 6**: Near-infrared SETI instrument schematic illustrating the optical-mechanical layout of the system with computers and control panel. All optical and mechanical components are purchased from commercial vendors.

## 3. COMMISSIONING AND OBSERVATIONAL PLANS

We plan to commission this instrument at Lick Observatory on the 1 meter Nickel telescope by the end of this calendar year. The primary observing program will be a targeted search of several thousand stars. With a field of view of ~4.4" on-sky this will encompass the seeing disk of unresolved sources. We intend to observe primarily remotely, which increases availability of observers, especially engaging students for educational purposes.

Our target selection is still being refined at this time. In 2012, our team used the previous optical SETI instrument to observe ~100 Kepler Objects of Interest (KOIs) that had more than one planet being confirmed, similar to Siemion et al.[31] recent observations using the Green Bank Telescope. Exoplanet statistics keep pointing to a larger and larger prevalence of planets around other stars, with one out of five FGK stars likely to harbor an exoplanet (Fressin et al.[32]). Kepler and Doppler surveys now imply that one out of ten sun-like stars (GK) has an earth-like planet (with an orbital period up to ~ 200 days; Petigura et al.[33]). Our original optical SETI search selected all FGKM stars within 200 light years. For near-infrared SETI we will define our sample at a range of distances out to 2 kpc, taking advantage of the decreased extinction. Our selection will likely be magnitude selected to optimize our sensitivity of our system.

We will also use the targeted search mode to monitor astrophysical sources of special interest. These may be objects which seem relatively likely as possible sources of short duration high energy pulses, such as pulsars, black holes, cataclysmic variables, gamma-ray bursters and active galactic nuclei. In addition, using the post-processing capabilities will allow us to look at astrophysical sources with a variety of time resolutions (μs to ns), and will provide a unique and powerful tool.

## 4. SUMMARY

We are poised to take advantage of a remarkable confluence of technological advancement with scientific opportunity. For the first time, very fast, wide bandwidth, high-gain, low noise near-infrared APDs are available and reasonably priced. We have presented the design of an innovative near-infrared SETI experiment that is affordable and easy to duplicate that will be one of the first probes of near-infrared SETI research. We make use of near-infrared (950 – 1650 nm) discrete amplification APDs from Amplification Technologies Inc. that have 80 μm active area with nano-second resolution. Our team measured the pulse height distributions of these discrete amplification APDs and determined the false coincidence rates for a near-infrared SETI experiment (see Maire et al., this conference). We present an economical opto-mechanical design that uses all off the shelf components. Our team plans to commission this instrument at the Lick Observatory Nickel (1m) telescope and begin remote observations for a targeted near-infrared SETI search.

An important benefit of this near-infrared SETI instrument will be introduction of a digital back-end that will greatly increase detection capabilities. A high bandwidth oscilloscope will allow us to improve sensitivity by implementing sophisticated and adaptable real time detection algorithms. The scope will either be set in a simple trigger mode for registering a pulse and coincidence between the two detectors, or we can set it to record the detector waveforms with one nanosecond timing resolution or higher. These recordings can later be examined in detail to help understand the nature of any detection or search for other underlying signals or be used at higher time samplings for astrophysical studies.

## ACKNOWLEDGEMENTS

We would like to thank the generosity of our anonymous donor for supporting this research and instrument; this individual's interest and enthusiastic support catalyzed these endeavors. We would like to extend our sincere thanks to Dr. Rafael Ben-Michael from Amplification Technologies Inc. for all of his help and patience for aiding our testing with the discrete amplification APDs. Everyone at Amplification Technologies Inc. was incredibly responsive and informative to all of our inquiries; it was a sincere pleasure working with this company. We would also like to thank Professor Hoi-Kwong Lo from the Department of Electrical and Computer Engineering at University of Toronto for lending testing equipment and for his useful discussions. Lastly, to the other extraterrestrial civilizations, may your lifetime be long and merry, and communicative.